\documentclass{acm_proc_article-sp}
\usepackage{listings}
\usepackage{tabularx}
\usepackage{url}
\usepackage{graphicx}
\usepackage[skip=1pt]{subcaption}
\usepackage{needspace}
\begin{document}

\title{Urban sidewalks: visualization and routing for individuals with limited mobility}

\numberofauthors{6}

\author{
\alignauthor
Nicholas Bolten \\
  \affaddr{\small{UW Electrical Engineering}}\\
  \email{\small{bolten@uw.edu}}
\alignauthor
Amirhossein Amini \\
  \affaddr{\small{UW Mechanical Engineering}}\\
  \email{\small{aminia@uw.edu}}
\alignauthor
Yun Hao \\
  \affaddr{\small{UW Department of Statistics}}\\
  \email{\small{yunhao@uw.edu}}
  \and
\alignauthor
Vaishnavi Ravichandran \\
  \affaddr{\small{University of Washington}}\\
  \email{\small{ravichandran.vaishu@gmail.com}}
\alignauthor
Andre Stephens \\
  \affaddr{\small{University of Washington}}\\
  \email{\small{andrestephens2000@gmail.com}}
\alignauthor
Anat Caspi \\
       \affaddr{\small{UW Computer Science \& Engineering}}\\
       \email{\small{caspian@cs.washington.edu}}
}
\date{8 October 2015}
\maketitle

\begin{abstract}
People with limited mobility in the U.S. (defined as having difficulty or inability to walk a quarter of a mile without help and without the use of special equipment) face a growing informational gap: while pedestrian routing algorithms are getting faster and more informative, planning a route with a wheeled device in urban centers is very difficult due to lack of integrated pertinent information regarding accessibility along the route. Moreover, reducing access to street-spaces translates to reduced access to other public information and services that are increasingly made available to the public along urban streets. To adequately plan a commute, a traveler with limited or wheeled mobility must know whether her path may be blocked by construction, whether the sidewalk would be too steep or rendered unusable due to poor conditions, whether the street can be crossed or  a highway is blocking the way, or whether there is a sidewalk at all. These details populate different datasets in many  modern municipalities, but they are not immediately available in a convenient, integrated format to be useful to people with limited mobility. Our project, AccessMap, in its first phase (v.1) overlayed the information that is most relevant to people with limited mobility on a map, enabling self-planning of routes. Here, we describe the next phase of the project: synthesizing commonly available open data (including streets, sidewalks, curb ramps, elevation data, and construction permit information) to generate a graph of paths to enable variable cost-function accessible routing.

\end{abstract}

\section{Introduction}

Our goal is to facilitate city-street traversal and routing for people with limited mobility. This is a particularly timely goal as public information and access to services is increasingly available for those physically present in city-streets. There is growing industry interest in using street-spaces to furnish public data to improve living conditions in cities and urban centers, for instance,
Inc.'s investment in Sidewalk Labs. It is of utmost importance to ensure that connectivity is truly equitable and that the information accessed by being 'connected' is enabling to all constituents in the same way, particularly as a whole industry grows around 'the connected city': equipping cities as connected public places where one can travel through any street and join free ultra high-speed Wi-Fi networks, access information about municipal services, and locate transit and wayfinding information.

The U.S. Center for Disease Control estimates that 7.3\% of adults living in the U.S. (or 17.2 million adults) are unable (or find it very difficult) to walk a quarter mile \cite{Blackwell2014}. What is challenging about planning a trip through city streets in a limited mobility situation? A typical street map is insufficient to determine whether a route has negotiable elevation changes, curb cuts, ramped passages, etc. Increasingly publicly available bodies of data that are relevant to anyone with limited mobility are not available in an easily consumable format. The initial phase of this project, AccessMap v1, featured integrated municipal sidewalk-related data with other publicly available information about the spatial landscape, all overlaid and visualized on a map\cite{AccessMap2015}.

Our current project is motivated by the need for automated routing in this scenario where a user's cost function may not prioritize the ordinary shortest path route, but express a function that combines multiple cost attributes when describing the cost of traversing a particular route. For example, a power wheelchair user may enter a majority preference for routes with curb cuts but still have minority preference for  overall shorter route distance. In AccessMap v1 we integrated and visualized a variety of information that had not previously been considered in routing systems. In order to implement an end-to-end routing solution that considers multiple attributes of the path and weights these attributes differently, we defined requirements for the next four development phases of AccessMap: a universal tool for preprocessing and integrating available street data from varied sources,  algorithms to create routable multi-attribute graph from the collected data, algorithms for multi-attribute routing, and finally, a simple user experience allowing users to explore the potentially large set of interesting routes resulting from a multi-attribute search.

Here, we describe our open source project which addresses two major challenges to achieving this level of cost function customization in routing: 1) data de-noising and integration and 2) processing required to generate a fully connected graph ready for multi-attribute routing. In the upcoming months we will complete our implementation of new algorithms for multi-attribute routing and the user experience frontend, with adequate performance evaluation methods for both.

We begin by describing our data integration and de-noising procedure. The raw municipal sidewalk data is noisy and comprises sidewalk segments (not connected in a graph), so no routing is possible. Our new beta open source software includes methods to preprocess municipal sidewalk data and to integrate additional data from external sources. Once the data is automatically cleaned and adjusted, we automated routable-graph construction producing a routing-ready multi-attribute graph (a graph which contains multiple attributes when assessing the cost of edge traversal in a routing algorithm). Routing in a multi-attribute environment may result in several, or many, interesting alternative routes. We are developing a graphical frontend to allow users to customize their cost functions, pose routing queries and interactively explore the routing query results.

\subsection*{Methods Overview}

\begin{figure}[tr]
\includegraphics[width=0.4\textwidth]{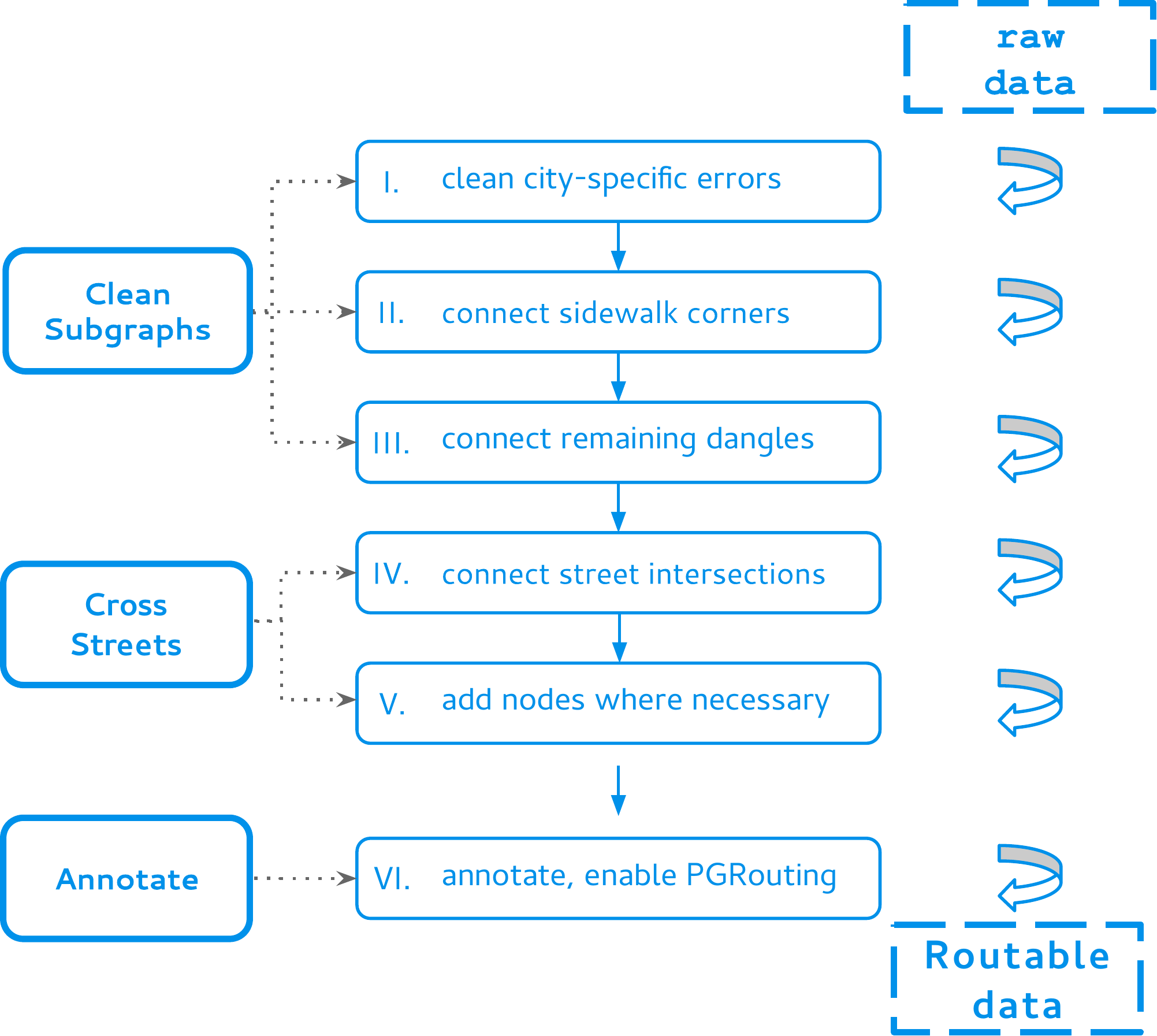}  
\caption{\bf{workflow for de-noising and processing multiple data types pertaining to sidewalk data}}
\label{fig:framework}  
\end{figure}

\begin{table*}[tb]
\begin{minipage}{0.9\textwidth}
	\centering
    \begin{tabular}{ l l l l }
\textit{Data Type} & \textit{Source} & \textit{Format} & \textit{Required De-noising}  \\ \hline
Sidewalks & Seattle DOT & Disconnected segments (LineStrings)  & Connectivity (sub-graphs)    \\
Streets & SDOT or KCM   & Connected segments (LineStrings) & No  \\
Curb Ramps & Sidewalks dataset  & Points extrapolated from sidewalks & Had to extrapolate location from data  \\
Elevations & National Elevation Dataset & Elevation in Meters & No  \\
Construction permits & data.seattle.gov  & LineStrings and points & Filtering to sidewalk impact, \\
 & & & active dates, assignment to \\
 & & & exact sidewalk location \\
\end{tabular}
  \caption{\bf{Table of data sources, format, and required de-noising.}}
\label{tab:datasettable}
\end{minipage}
\end{table*}

We have developed a platform-agnostic data processing framework (as shown in Figure \ref{fig:framework}) that generates an annotated, routable sidewalk graph from public municipal input data commonly available for cities and urban centers. With this data processing framework, only sidewalk segments, a street network, and curb ramp locations are required as inputs to generate a fully-connected and annotated sidewalk graph. We implemented our framework in Python, with the data de-noising and graph construction in SQL, leveraging the PostGIS geographical extension for PostgreSQL. Table \ref{tab:datasettable} shows the datasets we integrated for the purposes of this project.

\begin{figure}
  \centering
  \begin{subfigure}[t]{0.45\linewidth}
    \centering
    \includegraphics[width=\linewidth]{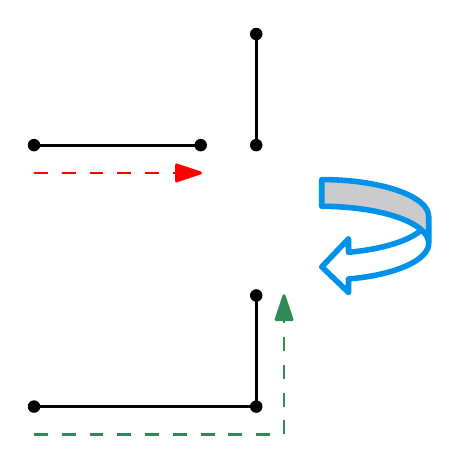}
    \caption{\bf{Connected sidewalk segments are routable}}
  \end{subfigure}
  ~
  \begin{subfigure}[t]{0.45\linewidth}
    \centering
    \includegraphics[width=\linewidth]{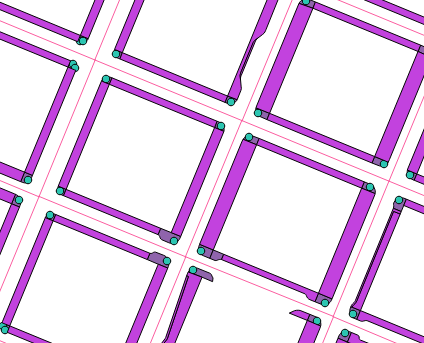}
    \caption{\bf{Sidewalks from orthoimagery}}
  \end{subfigure}

  \caption{\bf{(a) Noisy sidewalk data in the form of line segments. A routable path can be generated by connecting endpoints. (b) Sidewalks derived from orthoimagery data for the city of Portland. Such sidewalk polygons provide detailed information about sidewalk locations, but are not convenient for routing. In addition, they often have inaccuracies such as disconnections due to the process of extrapolating polygons from image data.}}
  \label{fig:linesvsareas}
\end{figure}

\begin{figure}[ht]
  \centering
  \begin{subfigure}[t]{0.45\columnwidth}
    \centering
    \includegraphics[width=\linewidth]{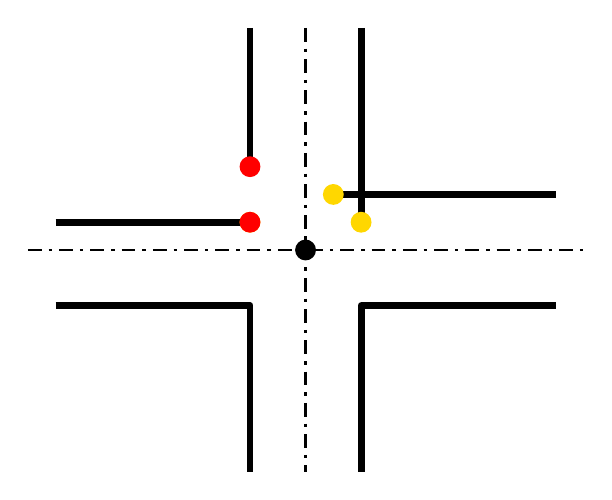}
    \caption{\bf{Corner classification}}
  \end{subfigure}
  ~
  \begin{subfigure}[t]{0.45\columnwidth}
    \centering
    \includegraphics[width=\linewidth]{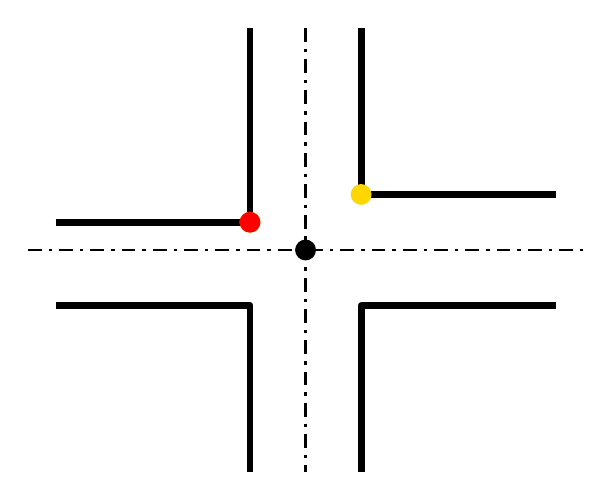}
    \caption{\bf{Corner connection}}
  \end{subfigure}

  \caption{\bf{Classification and connection of corners at intersections. Solid lines are sidewalks and dotted lines are street centerlines.}}
  \label{fig:corners}
\end{figure}

\begin{figure}
  \centering
  \begin{subfigure}[t]{0.45\linewidth}
    \centering
    \includegraphics[width=\linewidth]{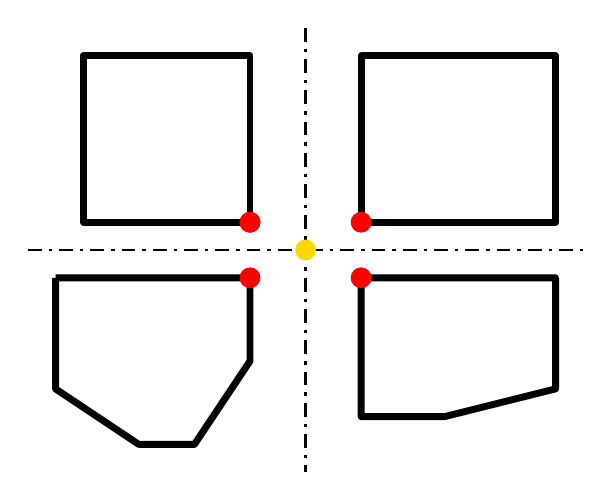}
    \caption{\bf{Crossing corner classification}}
  \end{subfigure}
  ~
  \begin{subfigure}[t]{0.45\linewidth}
    \centering
    \includegraphics[width=\linewidth]{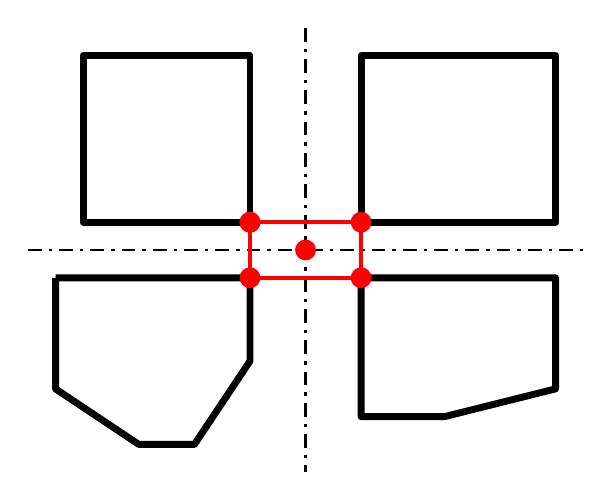}
    \caption{\bf{Connecting corners across streets}}
  \end{subfigure}

  \caption{\bf{(a) To connect sidewalks across streets, sidewalk corners near intersections are identified. (b) Corners are connected using a topological rule that enforces the crossing of a street to the nearest corner, or, if none is available, the closest point on the opposite sidewalk.}}
  \label{fig:crossing}
\end{figure}

\section{Data integration and \\de-noising procedure}

Open municipal data on sidewalks has not been optimized for routing, instead being generated from municipal asset management systems as non standardized disconnected line segments, with significant error in their coordinates. Alternatively, data may be generated from orthoimagery as polygonal descriptions of the sidewalk areas (Figure \ref{fig:linesvsareas} showing data types and desired outcome - a graph). Because routing requires a connected graph of line segments, data from these sources must be processed before sidewalk routing is possible. Our first urban center target, Seattle, WA, provides open data on sidewalks in the form of disconnected line segments with noisy coordinates. In an effort to build a universal tool, we looked beyond the scope of this initial urban area and projected how we may build upon this beta effort to de-noise and enhance data from other municipalities.

We integrated additional data to access other attributes that can inform our routing cost function. This required the added functionality of importing map data from standardized sources like Socrata for construction impacting the right-of-way and the National Elevation Dataset (NED) to calculate sidewalk slopes. These data required reformatting as well as appropriate feature selection (of informative annotation). An important reason for adequate feature selection is that several of the employed optimization criteria are not directly maintained in the maps. For example, we have information about elevation connected to the nodes (sidewalk ends) which have to be reassigned and post processed into edge attributes of the multi-attribute graph.

Considerable work is also required to consider the ways in which we served the proposed routes, favoring a clean, minimalist and inviting look that is both customizable and engaging to individuals with special mobility needs.

\subsection{City-specific error corrections}
To make our sidewalk de-noising strategy generalizable, we categorized the noise in the Seattle datasets as city-specific noise to be cleaned prior to entering the data processing tool's workflow, and city-agnostic noise that our tool would address. The primary Seattle-specific source of noise identified was a systematic error at 'T' intersections, where three streets meet and one of the angles between them is large. At these intersections, inaccurate and large gaps were systematically generated during sidewalk extrapolation from by the asset management system used by the Seattle Department of Transportation. We connected the sidewalks intersections by identifying 3-way intersections with angles between 170 and 190 degrees and connecting the appropriate sidewalks within 100 feet. We connected 88.2\% of the 4,352 'T' intersection sidewalks in the Seattle sidewalk data prior to entry into the data cleaning framework.

\enlargethispage{-10\baselineskip}

\section{automated routable-graph \\construction}

\subsection{Creating connected sidewalk graphs}

The initial step in our data processing framework is to connect sidewalk segments that should be connected in reality, but are disconnected in the data due to noise in the source data. The error in sidewalk locations can be large enough that endpoints erroneously appear on the wrong side of the street, appearing closer to neighboring sidewalk endpoints than to correct endpoints. As a result, proximity alone is insufficient to make a decision on whether two sidewalk segments should be connected. To address this, we developed a series of topological rules that first classify sidewalk endpoints by street blocks using street intersections before connecting sidewalk ends based on proximity (Figure \ref{fig:corners}). This approach relies only on prior knowledge of sidewalk location and street locations and can therefore be generalized to any city. Once connected, these sidewalks form subgraphs, typically representing the sidewalks connected around a block, but disconnected from one another (no across-street connections). We evaluated the coverage of our algorithms was evaluated by the number of modified sidewalks relative to the total number in the data set. Our cleaning and preliminary data analysis showed that we edited 86.9\% of the sidewalk corners at intersections and made 90.6\% of sidewalk subgraphs fully connected within their block.

\subsection{Connected subgraphs through \\intersections}

The next step in our data processing framework is to connect the subgraphs of sidewalks generated in the previous step to produce a highly-connected, routable graph. We based the decision to connect across streets on intersection corners and not other metadata, such as the existence of a crosswalk (Figure \ref{fig:crossing}), so all potential crossing locations are generated. Combining the sidewalk subgraph data with the street crossing data produces the final, fully-connected graph necessary for sidewalk routing. We evaluated the coverage of our crossing algorithm by the number of intersections that have at least one crossing, find that we generated crossings at 88.1\% of intersection corners.

\subsection{Annotating traversable edges}

\begin{figure}
  \centering
  \includegraphics[width=0.45\textwidth]{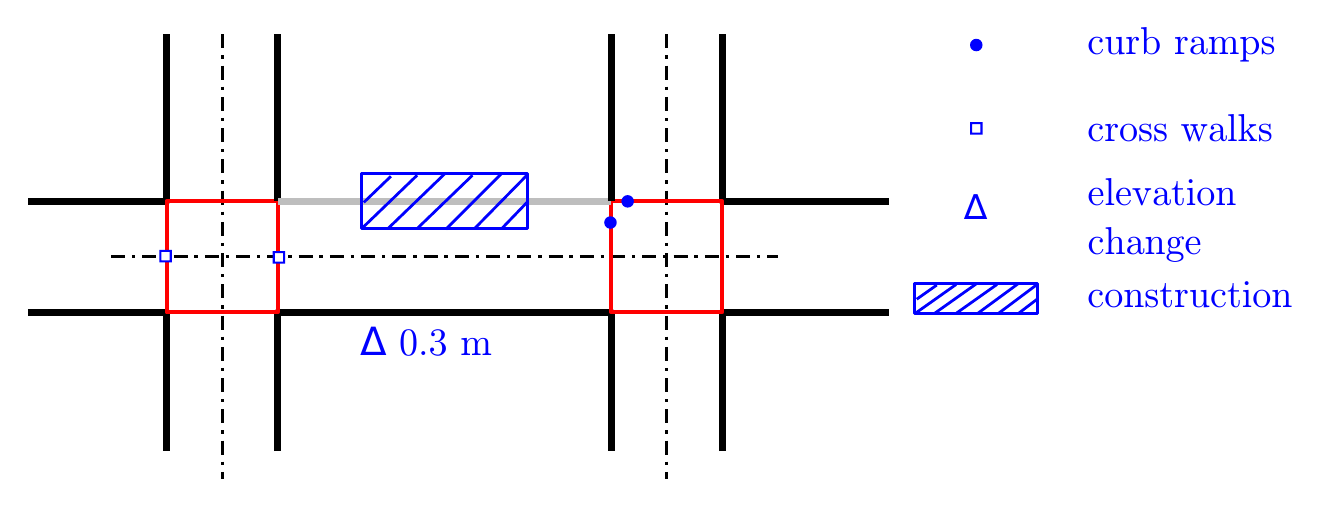}
  \caption{\bf{Annotation of sidewalk and street crossing segments by proximity. Point data, such as curb ramps, are annotated by proximity to corners or endpoints of segments. Line annotations, such as construction information, are annotated based on overall proximity to nearby segments.}}
  \label{fig:annotation}
\end{figure}

The final step in our graph construction process is to annotate the edges of the fully-connected graph, resulting in attribute labels for each sidewalk and street crossing that can then be used in a customizable routing cost function (Figure \ref{fig:annotation}). By default, elevation change information is assigned to every edge (sidewalks and street crossings) and the locations of curb ramps are assigned to street crossings. Because the cost function can use arbitrary attributes of edges in the graph, the labels added can be customized on a per-city basis depending on the availability of additional data sources. For example, the city of Seattle provides information on construction impacting the right-of-way through its open data portal, but this information may not be easily available for all cities. The annotated segments are then run through functions from the PGRouting package, which provides a table of nodes and layers from the segment data as well as a customizable SQL cost function for routing.

\section{Next development phases: \\multi-attribute routing and \\interactive route choices}

\subsection{Overview}

In our current implementation, producing routes depends on two software modules: (1) algorithms for finding optimal routes given user preferences and trip waypoint nodes (origin and destination(s)) and (2) a web API that translates desired trip waypoints (latitude-longitude pairs) into a format that the routing algorithms can use and returns step-by-step directions in a format readable by web mapping frameworks.

\subsection{Routing based on user preferences}

PGRouting provides optimized routing functions for a connected, annotated graph of the type produced by our data processing framework. Because a PGRouting cost function can include arbitrary attributes, functions, and scaling factors, we can tune its parameters for different user needs. For example, a wheelchair user may prefer curb ramps at street crossings, resulting in a high penalty for a street crossing lacking one or more curb ramps. Alternatively, a user of crutches may prioritize a shorter route over the existence of curb ramps, in which case curb ramps wouldn't factor into the cost function. We are in the process of testing what weightings and sub-functions best fit user preferences.

\subsection{Web API}

A wrapper around the SQL commands required to use PGRouting provides a human-usable interface for making and receiving trip requests. The SQL for requesting PGRouting routes must must specify start and end nodes via IDs generated automatically by the back-end and the return data is in a similar format of nodes and edges. Therefore, we have written a web API in Python that translates user requests for trips based on latitude-longitude waypoints into PGRouting commands, requests an optimal trip, and returns that trip in a standard JSON format including the route path and human-readable step-by-step instructions. We are in the process of extending the current AccessMap website to include this routing functionality in an easy to use format.

\section{Conclusions}
To address the challenge of improving street traversal for people with limited mobility, we have written an open source software package that takes common types of open data and produces an accessible routing database. To produce an accessible route, the database requires only a cost function and trip start and end points. The routing database is currently used in a simple user interface that optimizes for minimizing steep elevation traversal. In the near future, the routing database will be deployed within AccessMap and a native mobile application. Our longer-term goals are to use machine learning to enable adaptable de-noising and routable graph builds, to enhance the cost function capabilities, and to simplify the user experience without removing the expressivity of defining a weighted multi-attribute routing cost function.

User testing would identify preferable cost function parameters for different use cases, such as the best way to route a powered wheelchair user versus a manual wheelchair user, or for changing surface conditions, such as rain. Future directions include the application of this software framework to other cities and filling in gaps, such as parks or private areas, with OpenStreetMap data. Denver, CO and Savannah, GA have open data sets nearly identical to those of Seattle and could serve as initial testing cities. Many other cities have all of the data required but have orthoimagery sidewalk data and could be integrated into the workflow by approximating the orthoimagery sidewalk polygons with sidewalk segments.

\subsection*{Acknowledgements}
We acknowledge the work of Allie Deford, Veronika Sipeeva (AccessMap team), Alan Borning (UW CSE), Anthony Arendt, Jake Vanderplas (DSSG Data Scientists), and the support of the eScience Institute at the University of Washington.


%
\bibliographystyle{abbrv}
\bibliography{nyu-urbangis-sidewalks-refs}  
%
%
\end{document}